\documentclass[11pt,twoside,a4paper]{article}
\usepackage{amsmath}
\usepackage{amsfonts}
\usepackage{amssymb}
\usepackage{hyperref}
\usepackage{eqnarray}

\usepackage[page]{appendix}

\usepackage{multicol,lipsum,mathtools,enumitem} 

\usepackage{authblk}

\usepackage{todonotes}
\usepackage{leftidx}
\usepackage{tensor} 
\usepackage{xcolor} 
\usepackage{color} 
\usepackage{empheq} 
\usepackage[customcolors,shade]{hf-tikz}
\usepackage{kotex}
\usepackage{arydshln} 
\usepackage{multirow} 
\usepackage{enumitem} 
\usepackage{epsfig} 
\usepackage{tensor}

\def\Mp{\text{M}_{\text{p}}}
\def\bMp{\overline{\text{M}}_{\text{p}}}
\def\bphi{\bar{\phi}}
\def\bff{\bar{f}}
\def\tosc{t_{\text{osc}}}
\def\osc{\text{osc}}

\def\rh{\text{rh}}

\begin{document}

\baselineskip=16pt
\begin{titlepage}

\begin{center}

\vspace{0.5cm}

\large{\bf Modified Starobinsky Inflation}
\vspace*{5mm} \normalsize

{\bf Seokcheon Lee$^{\,\dagger}$} 

\smallskip
\medskip

$^\dagger${\it Research institute of natural science, Gyeongsang national university, 501 Jinju Daero, Jinju city, 52828, Korea}


\smallskip
\end{center}

\vskip0.6in

\centerline{\large\bf Abstract}
Starobinsky has suggested an inflation model which is obtained from the vacuum Einstein's equations modified by the one-loop corrections due to quantized matter fields. Although the one-loop gravitational action is not known for a general FRW background, it can be obtained in a de Sitter space to give $\Mp^2 R + \alpha R^2 + \beta R^2 \ln (R/M^2)$. Thus, one needs to investigate the inflationary behavior of this model compared to the Starobinsky model ({\it i.e.} $\beta = 0$). The coefficient $\alpha$ can be changed by varying the renormalization scale $M^2$ and $\beta$ is obtained from the quantum anomaly which is related to the numbers of quantum fields. It has been assumed that $\alpha \gg \beta$. We investigate the viable values of $\alpha$ and $\beta$ based on the CMB observation. We also scrutinize the reheating process in this model.      

\vspace*{2mm}

\end{titlepage}

\tableofcontents

\newpage

\section{Introduction}
In semi-classical gravity, one treats the background as classical, but with taking the back-reaction of the matter into account. In a curved space-time, even in the absence of classical matter or radiation, quantum fluctuations of matter fields give nontrivial contribution to the expectation value of the energy-momentum tensor being quantized in some state \cite{BD,9509057,9707062,BF,PT}. 
\begin{equation} \tensor{R}{_\mu_\nu} -\frac{1}{2} \tensor{g}{_\mu_\nu} R = 8 \pi G \langle \tensor{\hat{T}}{_\mu_\nu} \rangle \, ,  \label{1loop} \end{equation}
where $\langle \tensor{\hat{T}}{_\mu_\nu} \rangle$ includes $\infty^{4}$, $\infty^{2}$, and $\ln \infty$ divergences. The first and the second divergences can be eliminated by the renormalization of the cosmological constant and the Newtonian coupling constant, respectively. However, the logarithmic divergence can be eliminated only if one introduces an additional Lagrangian density of the form $\sqrt{-g} \left(c_1 R^2 + c_2 \tensor{R}{_\mu_\nu} \tensor{R}{^\mu^\nu} \right)$ \cite{Nariai73}. 

Among these quantum corrections to general relativity, if one includes only the quadratic terms in the Riemann tensors, then these terms are expressed as \cite{Mijic86}
\begin{equation} \alpha R^2 + \beta \tensor{R}{_\mu_\nu} \tensor{R}{^\mu^\nu} + \gamma \tensor{R}{_\mu_\nu_\lambda_\sigma} \tensor{R}{^\mu^\nu^\lambda^\sigma} = \epsilon R^2 + \xi \tensor{C}{_\mu_\nu_\lambda_\sigma} \tensor{C}{^\mu^\nu^\lambda^\sigma} + \eta \chi_{\text{E}} \label{quadraticR} \, ,\end{equation} where $C$ is the Weyl tensor and $\chi_{\text{E}}$ is the density of the Euler number for the manifold. If one considers the Robertson-Walker metric, then $C^2$-term vanishes because the metric is conformally flat. Thus, Starobinsky includes the curvature-squared correction to the Einstein-Hilbert action and this term is important for the early universe \cite{Star80}. At the early universe ({\it i.e.} when the curvatures are large), the curvature squared term is dominant than the Ricci scalar term and this correction term leads to an effective cosmological constant. Therefore, the early universe went through an inflationary de Sitter era without introducing an inflaton field.

These quantum corrections take a simple form in the case of free, massless, conformally invariant fields. The contributions which all fields make to the geometrical part of the vacuum to vacuum amplitude. These contributions originate in the vacuum polarization, which the background geometry induces, and give rise to nonobservable renormalizations as well as physically real radiative corrections \cite{DeWitt67}
\begin{equation} {\cal L}_{\text{vp}} = \beta R^2 \ln \left(\frac{R}{M^2} \right) \, . \label{Lvacuumpol} \end{equation}
Thus, when one regards the origin of the Starobinsky model which includes a quadratic Ricci scalar term in addition to the Hilbert-Einstein action, it is also natural to include the terms quadratic and logarithmic in the Ricci scalar. There have been investigation of the $f(R)$-gravity including the logarithmic correction for several aspects. The viable f(R)-gravity models replacing the dark energy have been considered \cite{0308176,0309062,0512109,07053199,07081482,08031081,09125474,13057290}. The logarithmic f(R) models are also considered to investigate the relativistic stars \cite{13077977}. Applying the logarithmic f(R)-gravity models for the early inflation have been scrutinized \cite{Vilenkin85,Baibosunov90,14047349,14061096,14100631,14116010,151209324}.  

Any regular model Universe must be non-singular for an infinitesimal perturbation of the metric 
\begin{equation} {\cal L}_{\text{reg}} = - \frac{M^{2}}{6} \left( \frac{R}{M^2} \right)^n \, . \label{Lregular} \end{equation}
The possible regular solutions of the gravitational equations in the presence of a nonlinear increment of the four curvature to the Lagrangian density of the gravitational field can be written as \cite{BGS71} 
\begin{equation} S_{\text{BGS}} = \frac{\Mp^2}{16 \pi} \int \sqrt{-g} \, d^4 x \left( R + M^2 f_{\text{c}}\left[ \frac{R}{M^2} \right] \right) \, , \label{BGSfR} \end{equation}
where $f_{\text{c}}(R/M^2)$ is a certain dimensionless function of the scalar curvature and $M$ is the characteristic mass. This work is motivated to obtain the bouncing Universe at $\tau = a t = 0$. 

As the minimal extension of the Starobinsky model, the one with the logarithmic correction term is given by \cite{Starobinsky79,Shore80,Vilenkin85,Baibosunov90,14047349}
\begin{align} S &= \frac{\Mp^2}{2} \int \sqrt{-g} \, d^4 x \left( R + \tilde{\alpha} R^2 + \tilde{\beta} R^2 \ln R \right) \nonumber \\ 
&=  \frac{\Mp^2}{2} \int \sqrt{-g} \, d^4 x R \left(1 + \alpha \frac{R}{M^2} + \beta \frac{R}{M^2} \ln \frac{R}{M^2} \right) \, , \label{SR2lnR} \end{align}
where $\alpha = \tilde{\alpha} + 2 \tilde{\beta} \ln M$ and $\beta = \tilde{\beta}$. This action is also naturally obtained when one considers the small correction in the Starobinsky model $ R + \alpha R^2 + R^{2+ \gamma}$ with $\gamma \ll 1$ as shown in \cite{14116010}. The inflationary behavior of this action was nicely investigated analytically by using the approximation \cite{14047349}. We improve the previous work numerically without using any approximation. We also investigate the reheating process in this model. Similar forms of the logarithmic model based on the stability and the viability of observation are also considered \cite{151209324,160508016}. Constraints on the model parameters given in this model should be obtained from the CMB observation \cite{150202114}. 

Higgs Starobinsky inflation models have been investigated to solve the large Higgs values in the early universe\cite{150908882,160502236,160906887,161104932,170106636,170107665,170505638,170700984}. In these models, non-minimal coupling of the Higgs field to the Ricci scalar induces the large quantum corrections. These approaches might explain both the dark matter and the inflation. 

In the usual inflation models, soon after the end of the inflation, the inflaton fields begin to oscillate around the minimum of their effective potentials, producing particles, which interact with each other to reach the thermal equilibrium at the reheating temperature, $T_{\text{rh}}$ \cite{10012600}. The reheating dynamics after the inflation induced by Starobinsky model have been investigated both in Einstein frame and  in Jordan frame \cite{09091737,12041472,12124466,160604346}.

The layout of this manuscript proceeds as follows. In the next section, we briefly review the equation of motions of the $R + \alpha R^2 + \beta R^2 \ln R$ inflation model both in Jordan frame and in Einstein frame. One needs to be careful when $\beta$ is negative. The approximate analysis of the inflationary behaviors was investigated in the reference \cite{14047349}. We scrutinize the exact inflation in this model for the different values of $\alpha$ and $\beta$ to be consistent with the Planck results. We investigate the oscillatory epoch and the reheating period of this model in section 3. Finally, we conclude in section 4.        

\section{$f(R)$ Inflation}
$f(R)$ inflation models are conventionally analyzed in the Einstein frame after the conformal transformation of the original action in the Jordan frame. In this section, we briefly review this process and provide the analytic relations formulae between the general form of $f(R)$ and the cosmological parameters obtained from the observations. We also investigate the specific form of $f(R)$ model obtained from the quantum correction of the matter to constrain its parameters.    

\subsection{General formalism}
The action for the general $f(R)$-gravity theories in 4 dimensional space-time can be rewritten as
\begin{equation} S \equiv \frac{1}{16 \pi G} \int \sqrt{-g} \, d^4 x f(R) \equiv \frac{\Mp^2}{2} \int \sqrt{-g} \, d^4 x M^2 \bar{f}(y) \, , \label{Sfx} \end{equation}
where $\Mp$ is the reduced Planck mass, $M$ is the characteristic mass scale, $y = R/M^2$ is the normalized dimensionless variable, and $\bar{f}$ is a dimensionless function. The above Jordan frame (JF) action is conformally transformed into the Einstein frame (EF) action by doing the Weyl transformation of the metric and the field redefinition
\begin{align} S_{\text{E}} &= \int  \sqrt{-g^{\text{E}} } \, d^4 x \, \left( \frac{\Mp^2 }{2} R_{\text{E}} - \frac{1}{2} \partial^{\mu} \phi \partial_{\mu} \phi - V \left( \phi \right) \right) \label{ESR2lnR} \, , \\
& \text{where}  \,\,\,\,\,\, g^{(\text{E})}_{\mu \nu} \equiv \frac{\partial f}{\partial R} g_{\mu\nu} \, , \label{gE} \\ 
\phi  &\equiv \Mp \bphi = \sqrt{\frac{3}{2}} \Mp \ln \left[ \, \frac{\partial f }{\partial R} \, \right]  = \sqrt{\frac{3}{2}} \Mp \ln \left[ \, \frac{\partial \bar{f} }{\partial y} \, \right] \equiv \sqrt{\frac{3}{2}} \Mp \ln \bar{f}_{,y} \left( y \right) \, , \label{bphi} \\ 
V \left( \phi \right) &\equiv \frac{\Mp^2}{2} \frac{R \left( \partial f / \partial R \right) - f}{ \left( \partial f / \partial R \right)^2}  = \frac{\Mp^2 M^2}{2} \frac{y \bar{f}_{,y} - \bar{f}}{ \bar{f}_{,y}^2} \, . \label{V} \end{align} 
The potential of the inflaton field, $\phi$ is obtained from the specific form of $f(R)$ from the theory. Thus, one can analyze the usual inflationary behavior of this model in the EF.  

One can obtain the slow-roll parameters as a function of $\bar{f}$ by using the Eq.(\ref{V})
\begin{align} \epsilon_{\text{V}} &\equiv \frac{\Mp^2}{2} \left( \frac{V_{,\phi}}{V} \right)^{2} = \frac{\left(-2 \bff+y \bff_{,y} \right)^2}{3 \left(\bff - y \bff_{,y}\right)^2}\label{epsilon} \, ,  \\
\eta_{\text{V}} &\equiv \Mp^2 \frac{V_{,\phi\,\phi}}{V} = -\frac{2 \left(\bff_{,y}^2-4 \bff \bff_{,yy} +y \bff_{,y} \bff_{,yy} \right)}{3 \left( \bff -y \bff_{,y} \right) \bff_{,yy}}\label{eta} \, , \\
\xi_{\text{V}}^{2} &\equiv 2 \Mp^4 \frac{V_{,\phi} V_{,\phi\phi\phi}}{V^2} = \frac{8 \left(2 \bff -y \bff_{,y}\right) \left(3\bff_{,y}^2 \bff_{,yy}^2-8 \bff \bff_{,yy}^3+y \bff_{,y} \bff_{,yy}^3+\bff_{,y}^3 \bff_{,yyy}\right)}{9 \left(\bff-y \bff_{,y}\right)^2 \bff_{,yy}^3}\label{xi} \, , 
\end{align}
where $V_{,\phi}$ means the derivative of $V$ with respect to the inflaton field, $\phi$. The second expressions in Eqs.(\ref{epsilon})-(\ref{xi}) are described by the dimensionless quantities and one can use these formulae without any mistake in units. The scalar spectral index $n_{s}$, the tensor-to-scalar ratio $r$, the running of the spectral index $d n_{s} /d \ln k$ and the scalar primordial amplitude $A_{s}$ are obtained from the above slow-roll parameters 
\begin{align} n_{s} &= 1 - 6 \epsilon_{\text{V}} + 2 \eta_{\text{V}}  \, , \label{ns} \\
r &= 16 \epsilon_{\text{V}} \label{r} \, , \\
\frac{d n_{s}}{d \ln k} &= 16 \epsilon_{\text{V}} \eta_{\text{V}} -24 \epsilon_{\text{V}}^2 -2 \xi_{\text{V}}^{2} \label{runns} \, , \\
A_{s} &= \frac{32}{75 \Mp^4} \frac{V}{\epsilon_{\text{V}}} \label{As} \, . \end{align} 

Thus, if one solves the evolution of $f(R)$, then one can obtain the model predictions for the inflationary evolution. In order to obtain the numerical analysis for these predictions, one needs to solve the field equation obtained from Eq.(\ref{ESR2lnR})
\begin{equation} \ddot{\phi} + 3H \dot{\phi} + V_{,\phi} = 0 \label{phiEq} \, , \end{equation}
where dot denotes the derivative with respect to the time, $t$. If one adopts the normalized field definition $\bMp \bphi = \phi$ and changes variable from $t$ to the e-folding number $N = \ln a$, then the above field equation becomes
\begin{equation} \bphi'' +  \left( 6 - \bphi^{'2} \right) \frac{\bphi'}{2} + \left( 6 - \bphi^{'2} \right) \frac{V_{,\bphi}}{2V} = 0 \label{bphiEq} \, . \end{equation}
The evolution of the given $f(R)$ as a function of $N$ is obtained when one inserts Eqs.(\ref{bphi})-(\ref{V}) into the Eq.(\ref{bphiEq}). We investigate the specific model of $f(R)$ in the following subsection. 

\subsection{$R + \alpha \frac{R^2}{M^2} + \beta \frac{R^2}{M^2} \ln \frac{R}{M^2}$ model}

\begin{figure*}[h]
\centering
    \includegraphics[width=1\linewidth]{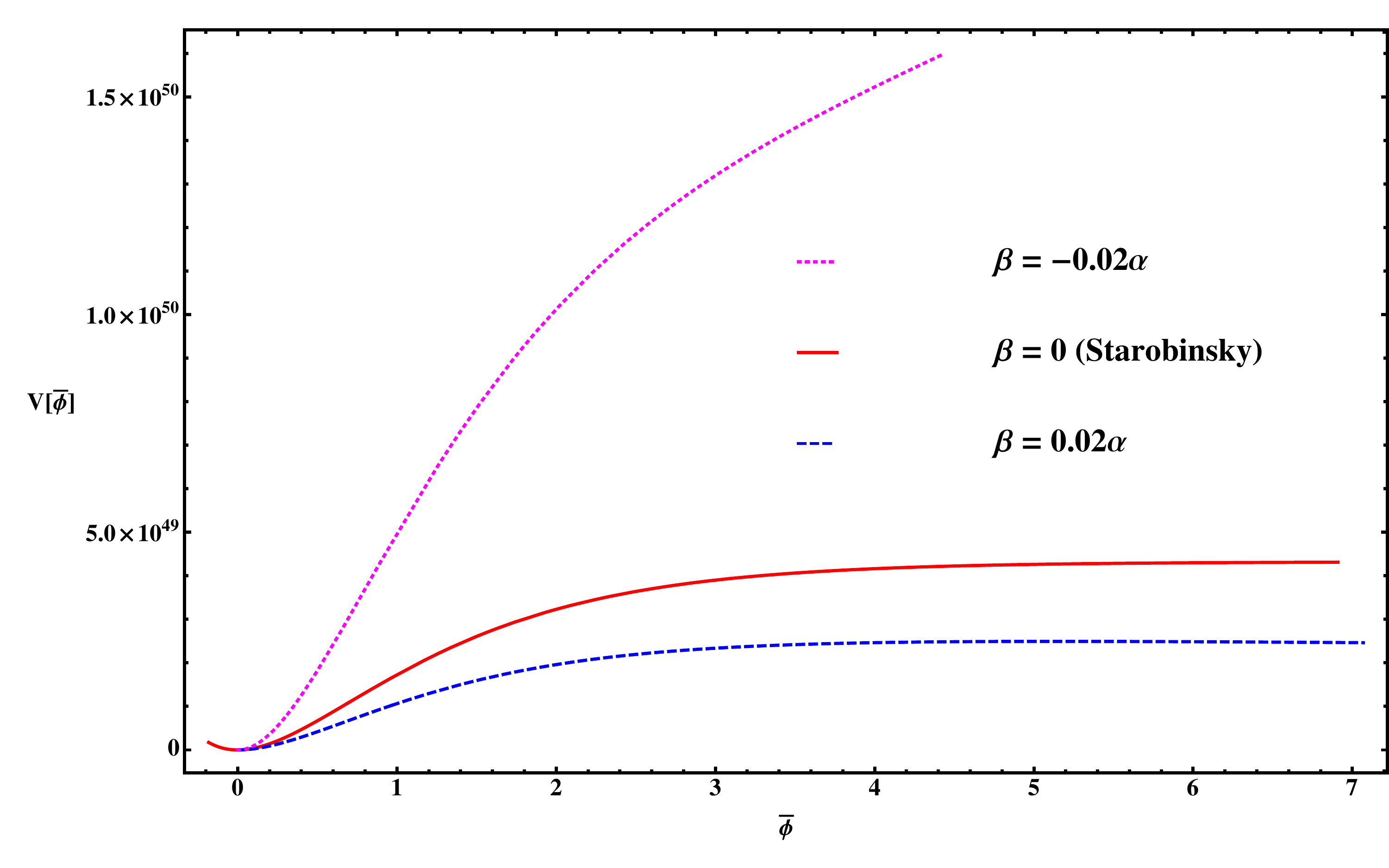} 
\caption{The form of potential for the different values of $\beta$. The solid, dotted, and dashed lines correspond $\beta = 0, -0.02$, and 0.02, respectively. The negative value of $\beta$ has a UV incomplete and the positive one has the runaway behavior.}
\label{fig-1}
\end{figure*}

As we mentioned in the introduction, the one loop corrections involving the matter fields in the curved space-time generally provide the correction terms with the quadratic and logarithmic in the Ricci scalar \cite{Nariai73,DeWitt67}. This action is given in Eq.(\ref{SR2lnR}), one can rewrite this as  
\begin{align} S &= \frac{\Mp^2}{2} \int \sqrt{-g} \, d^4 x R \left(1 + \alpha \frac{R}{M^2} + \beta \frac{R}{M^2} \ln \frac{R}{M^2} \right) \equiv \frac{\Mp^2}{2} \int \sqrt{-g} \, d^4 x M^2 \bar{f}(y) \nonumber \\ 
&= \frac{\bMp^2 M^2}{2} \int \sqrt{-g} \, d^4 x \, y \left(1 + \alpha y + \beta y \ln y \right) \, , \label{Sx2lnx} \end{align}
where $M$ is the characteristic mass scale can be tuned in order to obtain the proper inflationary behavior, $y = R/M^2$ is the normalized dimensionless variable (curvature), and $\bar{f}$ is a dimensionless function. For the given $f(R)$, one can obtain the analytic form of the Ricci scalar, and the scalar potential, $V(\phi)$ as nicely shown in the reference \cite{14047349}
\begin{align} R &= \frac{e^{\sqrt{\frac{2}{3}} \bphi} -1 }{2 \beta W_{k}(X)} = \frac{\bff_{,y} - 1}{2 \beta W_{k}(X)}  \label{RforR2} \,\,  , \\
X &\equiv \frac{e^{(2\alpha+\beta)/2\beta}}{2\beta} \left( e^{\sqrt{\frac{2}{3}} \bphi} -1 \right) = \frac{e^{(2\alpha+\beta)/2\beta}}{2\beta} \left( \bff_{,y} - 1 \right)   \label{XforR2} \,\, , \\
V(\bphi) &= \left(1 - e^{-\sqrt{\frac{2}{3}} \bphi} \right)^{2} \frac{1+2W_{k}(X)}{16 \beta W_{k}(X)^{2}} = \left( \frac{1 - \bff_{,y}}{\bff_{,y}} \right)^2  \frac{1+2W_{k}(X)}{16 \beta W_{k}(X)^{2}}  \label{VforR2} \, , \end{align} 
where $W_{k}$ is the ProductLog (Lambert function) of branch $k=0$ for the positive value of $\beta$ and $k=-1$ for the negative one. Even though there exist these exact analytic solutions, the numerical results for the dynamics of the system are more clear than the analytic ones due to the complexity of the Lambert function. Thus, we numerically solve the evolutions of the inflaton field. The behaviors of the inflaton potential for the different values of $\beta$ are shown in the Fig.\ref{fig-1}. $M = 6 \times 10^{12}$ GeV is used in this figure. The solid, dotted, and dashed lines depict the evolution of the inflaton potential when $\beta = 0, -0.02$, and 0.02, respectively. The requirement for the attractive gravity and ghost free graviton is given by $\partial f/ \partial R \, (\partial \bff/ \partial y) > 0$. In order to avoid a curvature singularity and Dolgov-Kaasaki instability, one also needs the condition $\partial^2 f/ \partial R^2 \, (\partial^2 \bff/ \partial y^2) > 0$ \cite{09091737}. These conditions provide the upper bound on the values of $\beta$ for the given value of $\alpha$ at each $R$. Thus, for the negative value of $\beta$, the model has the upper bound on the $\phi$ ({\it i.e. UV incomplete}). When $\beta = 0$, the model is identical to the Starobinsky model and show the plateau region to have the enough number of e-foldings. When the $\beta$ is positive, the potential becomes unstable with a runaway direction for the large values of $\phi$. In this case, the hilltop type model is realized when $\phi$ rolls towards the origin and $\beta$ value will be constrained in order to get the enough e-folding number.      

\begin{figure}[h]
\centering
\vspace{1cm}
\begin{tabular}{cc}
\epsfig{file=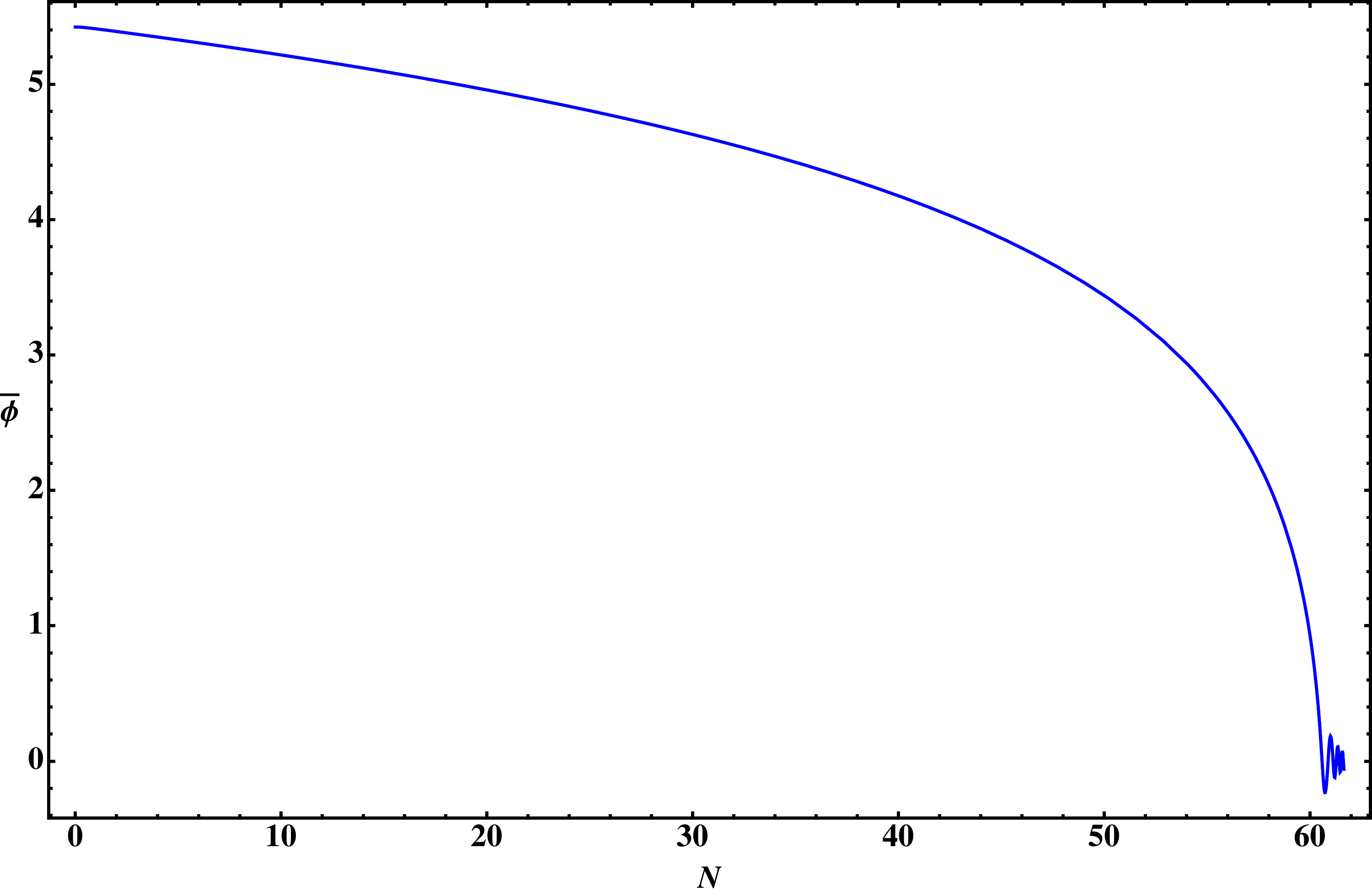,width=0.45\linewidth,clip=} &
\epsfig{file=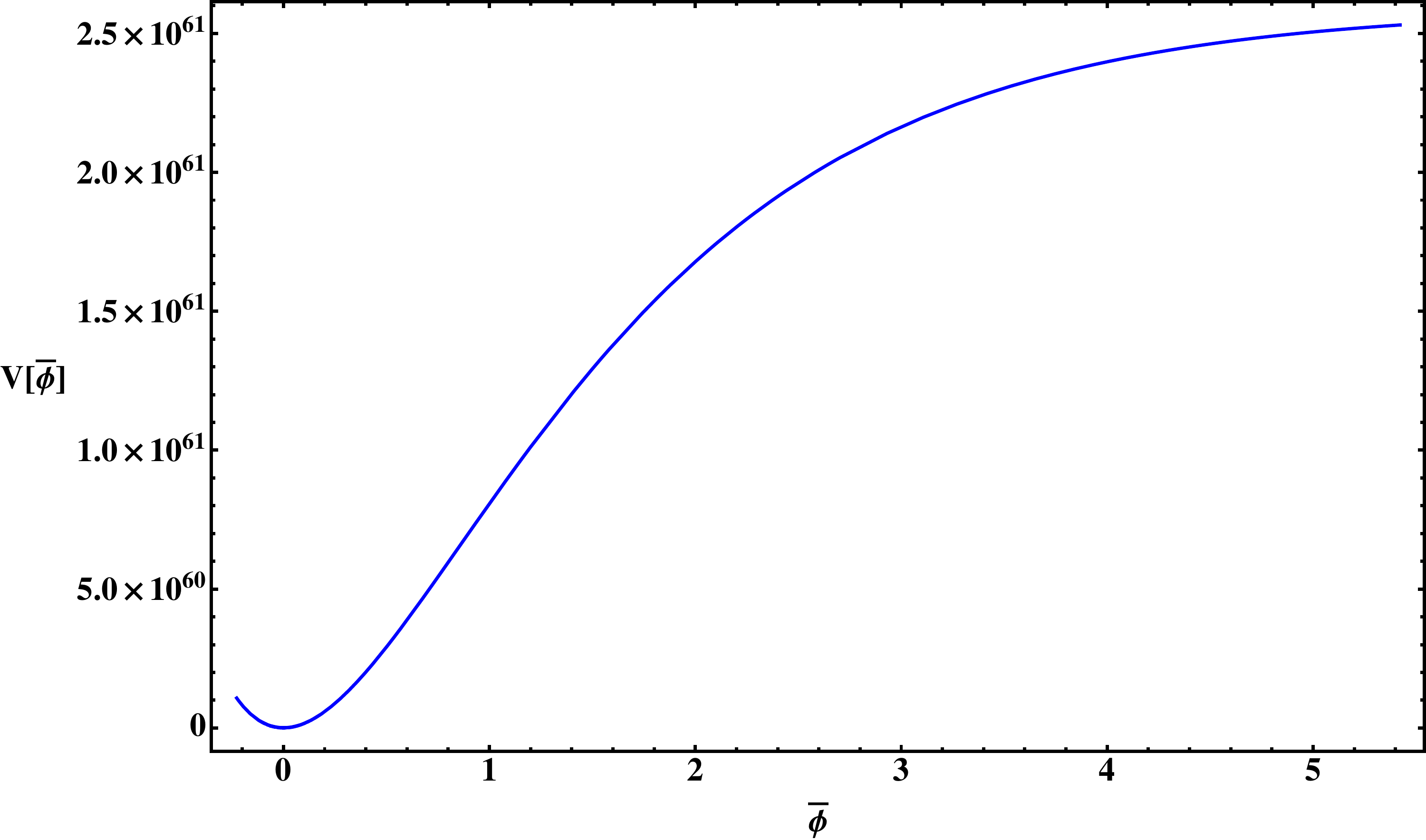,width=0.50\linewidth,clip=}
\end{tabular}
\vspace{-0.5cm}
\caption{The evolutions of $\phi$ and $V(\phi)$ during the inflation when $\beta =0$ to obtain $N_{\text{e-folindg}} = 60$. {\it left}) The evolution of the inflaton field as a function of the number of e-foldings. {\it right}) The behavior of the inflaton potential. } \label{fig-2}
\vspace{1cm}
\end{figure}

The detail evolution of the inflation model given in the action Eq.(\ref{Sx2lnx}) is investigated for the different values of coupling constant $\beta$. These results are obtained from Eqs.(\ref{bphi}), (\ref{V}), and (\ref{bphiEq}). We use $M = 6 \times 10^{12}$ GeV and $\alpha = 1$. In order to obtain the enough number of e-foldings, one needs to tune the initial values of the scalar field, $\bphi_{i}$. For an illustration, we show a typical evolutionary behavior of $\phi$ and $V(\phi)$ in Fig.\ref{fig-2}. In this figure, $\beta = 0$ and N-efolings are fixed at 60. In this model, $\bphi_{i}$ is chosen to be 41.36 in order to obtain $N_{\text{e-folding}} = 60$ ({\it i.e.} $\epsilon(N=60) =1$). In this case, the scalar field slowly rolls down through the flat plateau region and reaches to the minimum to oscillate. The oscillatory behavior of the $\bphi$ is shown in the left panel of the Fig.\ref{fig-2} and the slow-roll behaviors is represented on the right panel of the Fig.\ref{fig-2}.  

In tables.\ref{tab-1} and \ref{tab-2}, we summarize the values of the cosmological parameters obtained from the different values of $\beta$ when we fix the number of e-foldings. In table.\ref{tab-1}, we choose the e-folding number as 60 and vary the value of $\beta$ from -0.02 to 0.05. This range of $\beta$ provides the allowed regions for the cosmological parameters given in the reference \cite{150202114}. The smaller the values of $\beta$, the steeper the potential in the slow-roll plateau. Thus, one needs the larger initial field values for the smaller values of $\beta$ to obtain the same value of the e-folding number. The $\bphi_{i}$ varies from 75.2 to 16.4 for $-0.02 \leq \beta \leq 0.05$. One of slow-roll parameters, $\epsilon_{\text{V}}$ decreases as $\beta$ increases. However, as $\beta$ decreases, so does the other slow-roll parameter, $\eta_{\text{V}}$. The scalar spectral index, $n_{s}$ decreases as $\beta$ decreases. $n_{s}$ varies from 0.98 to 0.93 in the given range of $\beta$. The tensor-to-scalar ratio decreases from $7.4 \times 10^{-3}$ to $4.4 \times 10^{-4}$ when $\beta$ changes from -0.02 to 0.05. The magnitude of the running of the spectral index is maximum around Starobinsky model. The amplitude of the scalar primordial curvature perturbation increases as $\beta$ increases.   
\begin{table}
\centering
\caption{These are the cosmological parameters predicted from the models when one chooses $N_{\text{e-foldings}} = 60$.}
\label{tab-1}
\vspace{0.5cm}
\begin{tabular}{|c|c|c|c|c|c|c|c|}
\hline
\multirow{2}{*}{$\frac{\beta}{\alpha} (10^{-3})$}  & \multirow{2}{*}{$\bphi _i$} & \multirow{2}{*}{$n_s$} & $r$ & $\epsilon_{\text{V}}$ & $\eta_{\text{V}}$ & \multirow{2}{*}{$\frac{dn_{s}}{d \ln k} (10^{-4}) $} & $A_s$ \\
  &  & & $(10^{-3}) $ & $(10^{-4})$ & $(10^{-4}) $ & $$ & $(10^{-9})$ \\
\hline
$ -20$ & 75.19 & 0.980 & 7.41 & 4.63 & 0.76 & -5.22 & 0.78 \\
\hline
 $-10$ & 54.37 & 0.973 & 4.76 & 2.98 & 1.53 & -5.45 & 1.14 \\
\hline
$ -1$ & 42.42 & 0.968 & 3.25 & 2.03 & 2.42 & -5.51 & 1.61 \\
\hline
 0 & 41.36 & 0.967 & 3.12 & 1.95 & 2.54 & -5.51 & 1.67 \\
\hline
$1$ & 40.33 & 0.966 & 2.99 & 1.87 & 2.65 & -5.51 & 1.73 \\
\hline
 $10$ & 32.70 & 0.960 & 2.07 & 1.29 & 3.79 & -5.45 & 2.42 \\
\hline
 $20 $ & 26.65 & 0.953 & 1.39 & 0.87 & 5.31 & -5.30 & 3.48 \\
\hline
$ 50$ & 16.43 & 0.932 & 0.44 & 0.28 & 11.4 & -4.49 & 10.0 \\
\hline
\end{tabular}
\end{table}

The table.\ref{tab-2} shows the viable cosmological parameters when the e-folding number is 50. The $\bphi_{i}$ varies from 57.5 to 15.3 when $\beta$ changes  from -0.02 to 0.05. $\epsilon_{\text{V}}$ decreases from $5.8 \times 10^{-4}$ to $5.5 \times 10^{-5}$ for the given range of $\beta$. $\eta_{\text{V}}$ increases from $1.33 \times 10^{-4}$ to $1.33 \times 10^{-3}$ for the same interval of $\beta$. The scalar spectral index, $n_{s}$ decreases as $\beta$ decreases. $n_{s}$ varies from 0.97 to 0.93 in the given range of $\beta$. The tensor-to-scalar ratio decreases from $9.2 \times 10^{-3}$ to $8.9 \times 10^{-4}$ when $\beta$ changes from -0.02 to 0.05. The magnitude of the running of the spectral index is maximum around Starobinsky model. The amplitude of the scalar primordial curvature perturbation increases from $6.2 \times 10^{-10}$ to $5.0 \times 10^{-9}$ as $\beta$ increases.   
\begin{table}
\centering
\caption{These are the cosmological parameters predicted from the models for $N_{\text{e-foldings}} = 50$.}
\label{tab-2}
\vspace{0.5cm}
\begin{tabular}{|c|c|c|c|c|c|c|c|}
\hline
\multirow{2}{*}{$\frac{\beta}{\alpha} (10^{-3})$}  & \multirow{2}{*}{$\bphi _i$} & \multirow{2}{*}{$n_s$} & $r$ & $\epsilon_{\text{V}}$ & $\eta_{\text{V}}$ & \multirow{2}{*}{$\frac{dn_{s}}{d \ln k} (10^{-4}) $} & $A_s$ \\
  &  & & $(10^{-3}) $ & $(10^{-4})$ & $(10^{-4}) $ & $$ & $(10^{-9})$ \\
\hline
$ -20$ & 57.45 & 0.973 & 9.23 & 5.77 & 1.33 & -7.67 & 0.62 \\
\hline
$-10$ & 43.76 & 0.967 & 6.35 & 3.97 & 2.35 & -7.88 & 0.85 \\
\hline
$-1$ & 35.41 & 0.961 & 4.60 & 2.88 & 3.46 & -7.95 & 1.13 \\
\hline
 0 & 34.64 & 0.960 & 4.44 & 2.78 & 3.60 & -7.95 & 1.17 \\
\hline
$1$ & 33.90 & 0.960 & 4.29 & 2.68 & 3.73 & -7.95 & 1.20 \\
\hline
$10$ & 28.27 & 0.954 & 3.15 & 1.97 & 5.08 & -7.89 & 1.58 \\
\hline
$20$ & 23.63 & 0.947 & 2.27 & 1.42 & 6.79 & -7.73 & 2.13 \\
\hline
$50$ & 15.31 & 0.927 & 0.89 & 0.55 & 13.3 & -6.90 & 5.00 \\
\hline
\end{tabular}
\end{table}
 
Thus, even though the logarithmic corrections to the Starobinsky model is natural if one accepts the fact that $f(R)$-gravity is originated from the quantum corrections of the matter in the curved space-time, the correction is strongly constrained in order to satisfy the observation. In semi-classical theory, one can obtain the $R^2$-term from the local action and $R^2 \ln R$-term from the non-local action. The coupling constant $\alpha$ can be changed by varying the renormalization scale $M$, while the coefficient $\beta$ is related to the trace anomaly. $\beta$-value can be determined by the numbers of quantum fields for the different spins. And usually $\alpha \gg \beta$ is satisfied \cite{Vilenkin85}. Thus, if one can obtain the constrain of $\beta$ accurately, then one might be able to obtain the number of fields during the early universe. Even though,$\beta$ should be less than 1\% compared to $\alpha$, its effect on the cosmological parameters can be distinguishable if the accuracies of observations reach to percentage level.   

\section{Reheating}
The basic reheating process of the Universe at the end of the inflation is that the oscillating inflaton fields produce radiations via the tree-level decay of inflaton particles into relativistic species. Reheating is terminated when the rate of expansion of the Universe becomes smaller than the total decay rate of the inflaton field into new fields. Even though this perturbative reheating process is not the full story, it is good enough to estimate the reheating epoch. Thus, we show the oscillatory epoch and reheating period based on the given model. The reheating process and reheating temperature based on the Starobinsky model have been investigated in the literature \cite{09091737,12041472,12124466,160604346}. We probe the reheating process for the $R^2 \ln R$-correction model and investigate any difference compared to those of the Starobinsky model.    

\subsection{Oscillation}
At the end of the inflation, the scalar fields reach to the minimum of the potential and start to oscillate around it. In order to describe this period properly,  one needs to describe this epoch semi-analytically. The field equation of the general $f(R)$-gravity theories is obtained from the action (\ref{Sfx})
\begin{equation} f'(R) \tensor{R}{_\mu_\nu} - \frac{1}{2} f(R) \tensor{g}{_\mu_\nu} + \left[ \tensor{g}{_\mu_\nu} - \nabla_{\mu} \nabla_{\nu} \right] f'(R) = \frac{1}{\Mp^2} \tensor{T}{_\mu_\nu} \, , \label{fEq} \end{equation} 
where prime means the derivatives with respect to $R$. For the homogeneous and isotropic background, one can obtain two field equations from the above Eq. (\ref{fEq})
\begin{align} \dot{R} &= - \frac{f'}{f''} H + \frac{1}{6H} \frac{f' R - f}{f''} + \frac{1}{f''H} \frac{\rho}{3 \Mp^2} \, , \label{dotR} \\
\ddot{R} &= - 3H \dot{R} - \frac{f'''}{f''} \dot{R}^2 + \frac{ R f' - 2f}{3 f''} + \frac{1}{f''} \frac{\rho - 3P}{3 \Mp^2} \, , \label{ddotR} 
\end{align}
where $\rho$ and $P$ are the energy density and the pressure of some species, respectively.  If one adopts the specific form of $f(R)$ given by Eq.(\ref{SR2lnR}), then one obtains 
\begin{align} & \dot{R} + \left(\frac{M^2 }{2\alpha+3\beta+2\beta \ln x} \right) H + \left(\frac{2\alpha+\beta+2\beta \ln x }{2\alpha+3\beta+2\beta \ln x} \right) H R \nonumber \\ 
&- \left(\frac{ \alpha+\beta+\beta \ln x }{2\alpha+3\beta+2\beta \ln x} \right) \frac{R^2}{6H} = \left(\frac{M^2}{2\alpha+3\beta+2\beta \ln x} \right)  \frac{\rho}{3 \bMp^2 H} \, , \label{dotR2} \\
& \ddot{R} + 3H \dot{R} + \left(\frac{2\beta}{2\alpha+3\beta+2\beta \ln x} \right)\frac{\dot{R}^2}{R} + \frac{M^2}{3} \left(\frac{1-\beta x}{2\alpha+3\beta+2\beta \ln x} \right) R \nonumber \\ 
&= \left(\frac{M^2}{2\alpha+3\beta+2\beta \ln x} \right)\frac{\rho - 3P}{3 \bMp^2} \, . \label{ddotR2} 
\end{align}
One can use the relation between the Ricci scalar and the Hubble parameter, $R = 6 \left( \dot{H} + 2 H^2 \right)$ to investigate the evolution of $H$. Then, the equation (\ref{dotR}) is rewritten as
\begin{align} & \ddot{H} - \left( \frac{\alpha+\beta+\beta \ln x}{2\alpha+3\beta+2\beta \ln x} \right) \frac{\dot{H}^2}{H} + 3 H \dot{H} + \frac{M^2}{6 \left( 2\alpha+3\beta+2\beta \ln x \right)} H \ \nonumber \\
& - \frac{2\beta}{2\alpha+3\beta+2\beta \ln x } H^3 =  \left(\frac{M^2}{2\alpha+3\beta+2\beta \ln x} \right) \frac{\rho}{3 \bMp^2 H} \, . \label{dotR3} \end{align} 
Now one can scrutinize the behaviors of $R$ and $H$ from Eqs.(\ref{dotR2})-(\ref{dotR3}). First, one can puts $\dot{R} = 0$ and $\rho = 0$ in the slow-roll region. Then one obtains 
\begin{align} & R_{i} \simeq 3 b H_{i}^2 \left( 1 + \sqrt{1 + \frac{2c}{3b^2 H_{i}^2}} \, \right) \label{Ri} \, , \\
& \text{where} \,\, b \equiv \frac{2\alpha+\beta+2\beta \ln x}{\alpha+\beta+\beta\ln x} \,\, , \,\, c \equiv \frac{M^2}{\alpha+\beta+\beta\ln x} \, . \label{bc} 
 \end{align}    
As one expects $R_{i} \sim 12 H_{i}^2$ in this region, because $b \sim 2$ when $\alpha \gg \beta$. Even though $H$ is almost constant, it decreases almost linearly in time with a small slope. This can be obtained from Eq.(\ref{dotR3})
\begin{align} \dot{H} &\simeq - \frac{M^2}{18 \left(2\alpha+3\beta+2\beta \ln x \right)} + \frac{2\beta}{3 \left(2\alpha+3\beta+2\beta \ln x \right) } H^2 \nonumber \\ 
& \simeq - \frac{M^2}{18 \left(2\alpha+3\beta+2\beta \ln x \right)} \label{SRdotH} \, , \end{align}
where we use the approximation $\beta \ll 1$ in the second equality. In order to obtain the finite period of inflation, $\left(2\alpha+3\beta+2\beta \ln x \right)$ should be positive. One also obtains the Hubble parameter from the Eq.(\ref{SRdotH}) 
\begin{equation} H(t) \simeq H_{i} - \frac{M^2}{18 \left(2\alpha+3\beta+2\beta \ln x \right)} \left( t - t_{i} \right) \, . \label{Hit} \end{equation}
From this, one can estimate the expansion of the scale factor of the Universe during this region 
\begin{equation} a(t_{\text{end}}) \simeq a_{i} \exp \left[ \frac{9 \left( 2\alpha + 3\beta + 2\beta \ln x_{i} \right) H_{i}^2}{M^2} \right] \label{SRa} \, , \end{equation}  
where we use the fact that $\ln x$ is almost constant in this region. Thus, one can constrain the $M^2$ to satisfy the e-foling number for the given values of $\alpha$ and $\beta$. We already do this numerically and do not need to probe any detail in this section. 
Second, we consider the oscillation period. During the slow-roll period, $\Bigl|\frac{1}{2} \frac{\dot{H}^2}{H} \Bigr| \ll \Bigl|3 \dot{H} H \Bigr|$ is satisfied. However, the magnitudes of these two terms become comparable as $H$ decreases and becomes small. And oscillatory phase is followed with $\Bigl|3 \dot{H} H \Bigr| \simeq 0$. 
\begin{equation} \ddot{H} - \left( \frac{\alpha+\beta+\beta \ln x}{2\alpha+3\beta+2\beta \ln x} \right) \frac{\dot{H}^2}{H} + \frac{M^2}{6 \left( 2\alpha+3\beta+2\beta \ln x \right)} H + \frac{2\beta}{2\alpha+3\beta+2\beta \ln x } H^3 =  0 \, . \label{dotR3osc} \end{equation}
In order to obtain the oscillation period, $H^3$-term should be negligible ({\it i.e.} $\beta \simeq 0$). With this condition, the solution for Eq.(\ref{dotR3osc}) is given by
 \begin{align} & H(t) = c_{1} \times \cos^{\frac{M^2}{6 \omega^2}} \left[ \omega (t - t_{\text{osc}}) \right] \, , \nonumber \\ 
 & \text{where} \,\, \omega = \frac{M \sqrt{\alpha+2\beta+\beta \ln x}}{\sqrt{6} (2\alpha+3\beta+2\beta \ln x)} \label{Hosc} \,\,  \text{and} \,\, t_{\text{osc}} =  6 c_{2} \left( 2\alpha+3\beta+2\beta \ln x \right) \, . \end{align}
$c_1$ and $c_2$ are integral constants. $t_{\text{osc}}$ is the time when the oscillation period begins with $\Bigl|\frac{1}{2} \frac{\dot{H}^2}{H} \Bigr| = \Bigl|3 \dot{H} H \Bigr|$. One can estimate the order of the frequency, $\omega$ at this epoch. ${\cal O} (\omega) =  {\cal O} (M/\hbar) \sim 10^{37}$ Hz.   
We point out the oscillatory behavior before derive the approximate solution. The oscillation is powered by $\frac{M^2}{6\omega^2}$ and thus it should be an integer. This means $\beta \simeq 0$. Also, the oscillation should be damped and thus, one needs to replace $c_{1}$ with some function $g(t)$ in Eq.(\ref{Hosc}). Now, one needs to specify the $g(t)$ in the $\alpha \gg \beta$ limit. For this we compare the $H(t)$s in both before and during the oscillation region. 
	\begin{equation} \label{Htiosc}
		H(t) = \begin{dcases} H_{i} - \dfrac{2\omega^2}{3} \left(t-t_{i} \right) = \dfrac{\omega}{3} \left( 1 - 2 \omega \left( t - \tosc \right) \right)  &, \,\,  t_{i} < t \leq  \tosc \\
						    \dfrac{\cos^{2} \left[\omega \left(t - \tosc \right) \right] }{\left(\frac{15}{2\omega }+6(t-\tosc)-\frac{9}{2\omega } \cos [2\omega (t-\tosc)]\right)}  & , \,\, \tosc \leq t \, ,
		            \end{dcases}
	\end{equation} 
where $\tosc = t_{i} + (3/2\omega^2) H_{i} - (1/2\omega)$. We show the details to obtain these solutions in the appendix. Thus, after the long inflationary plateau, $H(t)$ decreases linearly and reaches to the oscillation period. The approximate solutions for the scale factor $a$ are obtained from $H$s in Eqs.(\ref{Htiosc}) 
	\begin{equation} \label{attiosc} a(t) = \begin{dcases} a_{i} e^{H_{i}\left(t - t_{i}\right) - (\omega^2/3) \left( t -t_{i} \right)^2} &, \,\, t_{i} < t \leq t_{\text{osc}}  \\
			 a_{\osc} \left( 1 + \dfrac{\omega \left(t - \tosc \right)}{4} \right)^{2/3}  &, \,\, \tosc \leq t \, , 
			 \end{dcases}
	\end{equation}
where one can obtain the $a(t)$ in the oscillatory phase by integrating the $H$ averaged over a few cycles. 
During the corresponding periods, the Ricci scalar evolves as
\begin{equation} \label{Rtiosc} 
	R(t) = \begin{dcases} -4 \omega (\omega -1) - 8\omega ^2 \left( t-t_{i} \right) &, \,\, t_{i} < t \leq t_{\text{osc}} \, , \\ 
                                              \frac{-4 \omega  \cos \left[\left(t-t_{\text{osc}}\right) \omega \right]}{\left(5+4 \omega \left( t - t_{\text{osc}} \right) -3 \cos \left[2 \left(t-t_{\text{osc}}\right) \omega \right]\right){}^2} \times & \\ 
                                              \Biggl( \Bigl(-7+\left(4-8 t+8 t_{\text{osc}}\right) \omega \Bigr) \cos \left[\left(t-t_{\text{osc}}\right) \omega \right] & \\ 
                                              + 3 \cos \left[3 \left(t-t_{\text{osc}}\right) \omega \right]+8 \omega  \Bigl( 2+\left(t -t_{\text{osc}}\right) \omega \Bigr) \sin [(t-\tosc) \omega ] \Biggr) &, \,\, \tosc \leq t  \, . \end{dcases}
\end{equation}

\begin{figure}[h]
\centering
\vspace{1cm}
\begin{tabular}{cc}
\epsfig{file=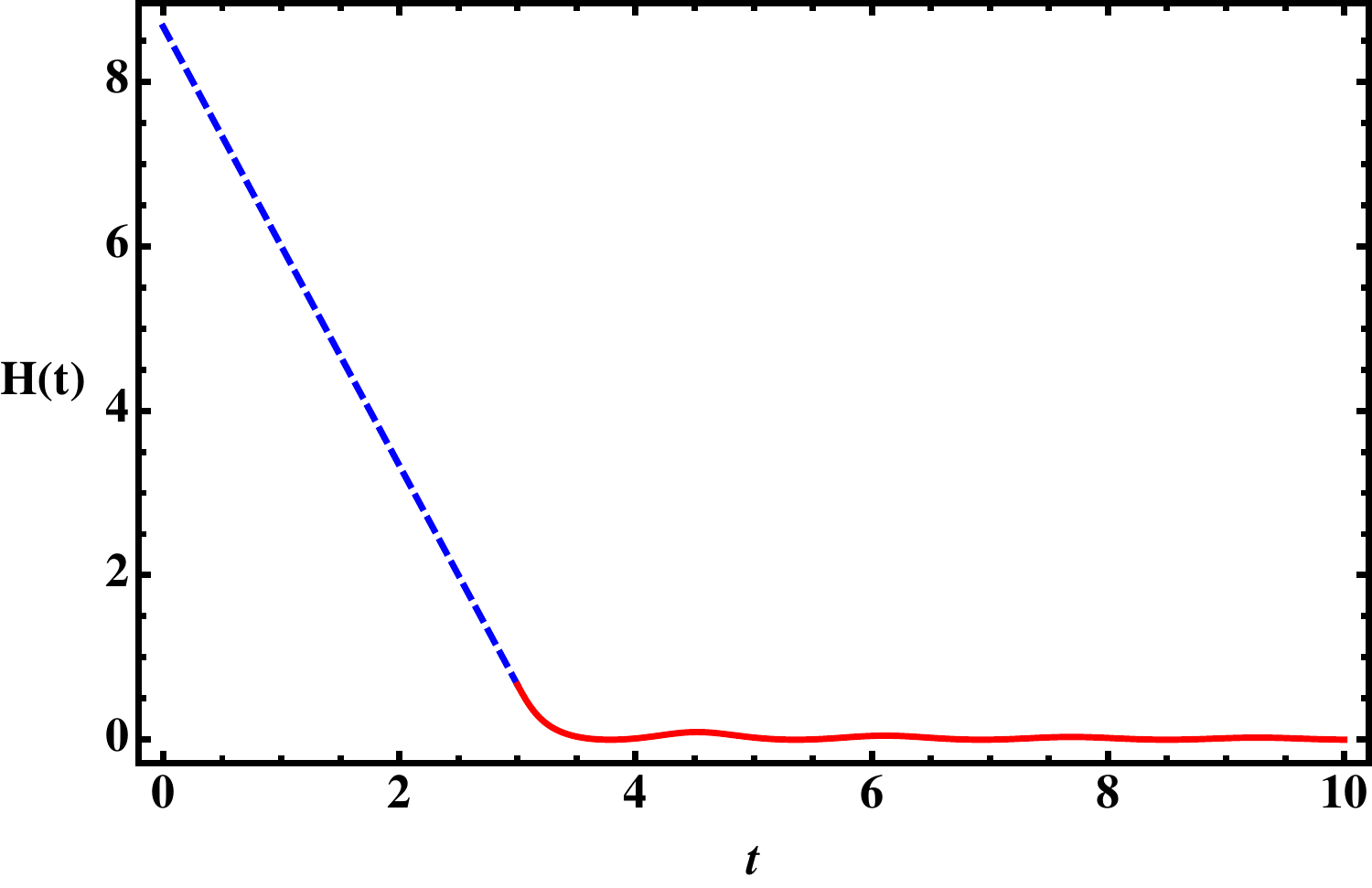,width=0.5\linewidth,clip=} &
\epsfig{file=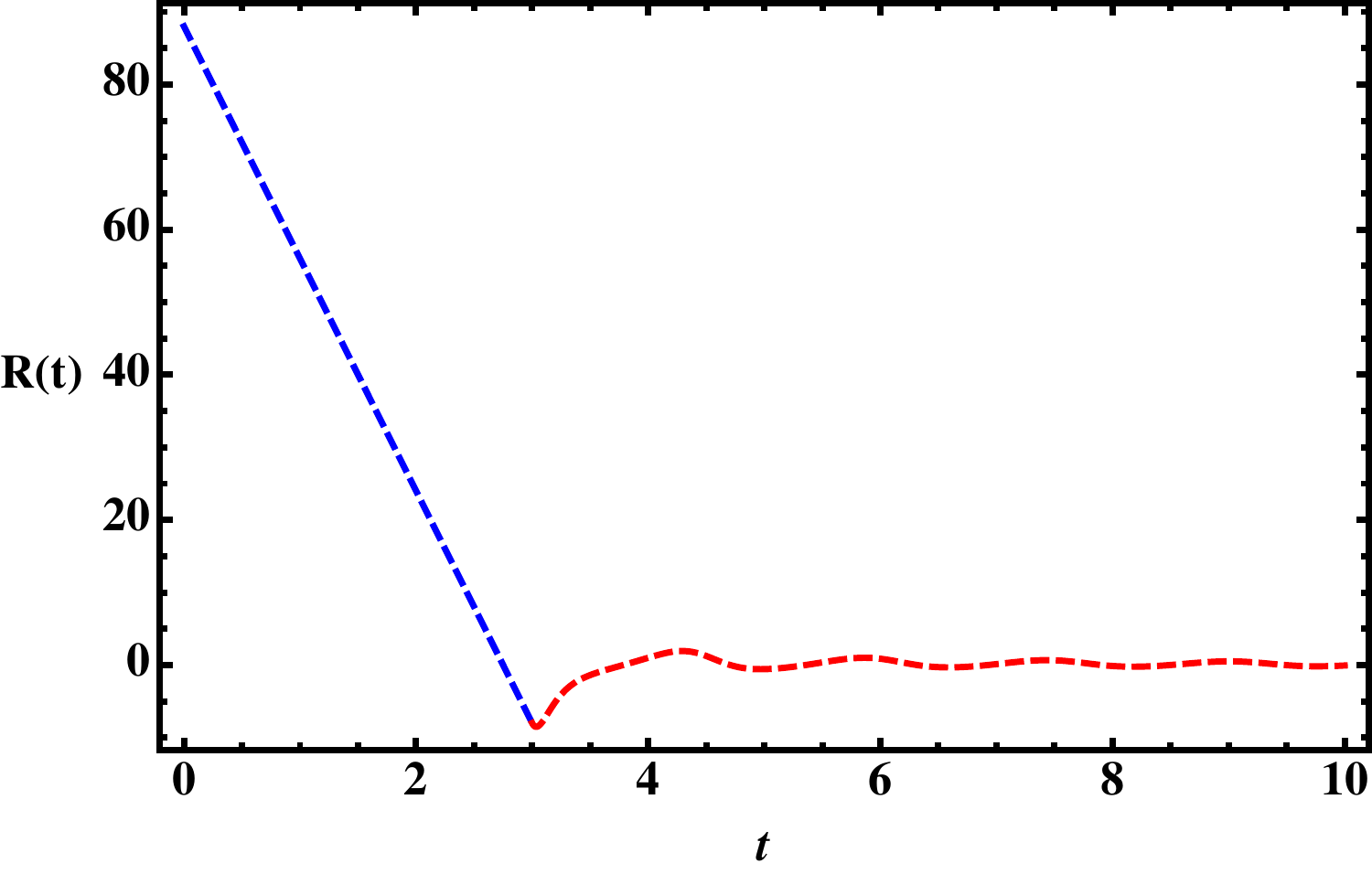,width=0.5\linewidth,clip=}
\end{tabular}
\vspace{-0.5cm}
\caption{The evolutions of $H(t)$ and $R(t)$ after inflation epoch when $M = 6 \times 10^{12}$ GeV and $\alpha = 1$ . Oscillation period is followed after a linearly decreasing period. {\it left} ) The evolution of H(t). {\it right}) The evolution of the Ricci scalar.} \label{fig-3}
\vspace{1cm}
\end{figure}
We show the evolution of $H$ and $R$ in Fig.\ref{fig-2}.  One should note that the shape and size of the oscillatory phase depend only on the $M$ and the $\alpha$ because one should use $\alpha \gg \beta$ in order to obtain the oscillatory behavior on the Hubble parameter as shown in the above.
 
\subsection{Reheating}
\begin{figure}[h]
\centering
\vspace{1cm}
\begin{tabular}{cc}
\epsfig{file=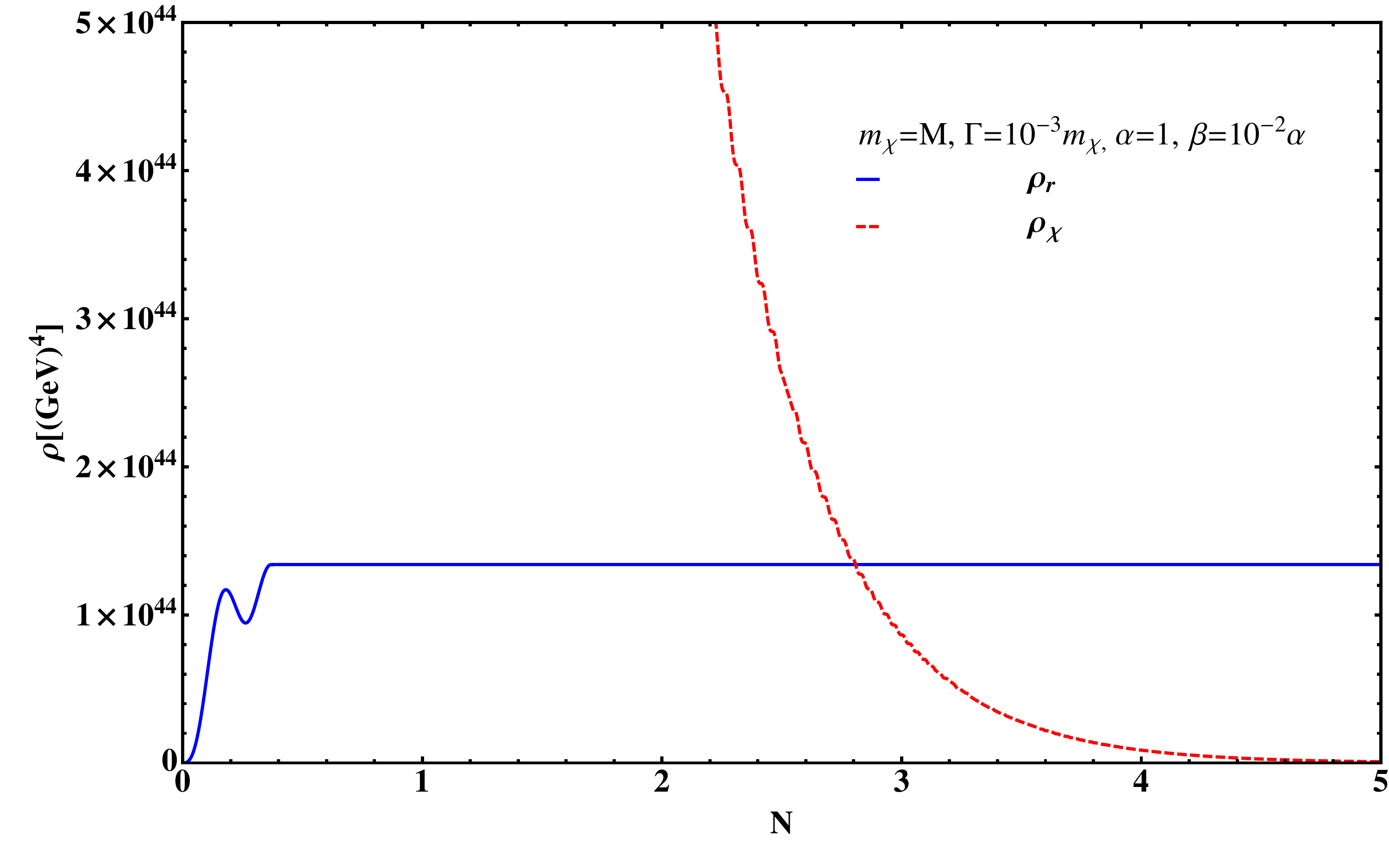,width=0.5\linewidth,clip=} &
\epsfig{file=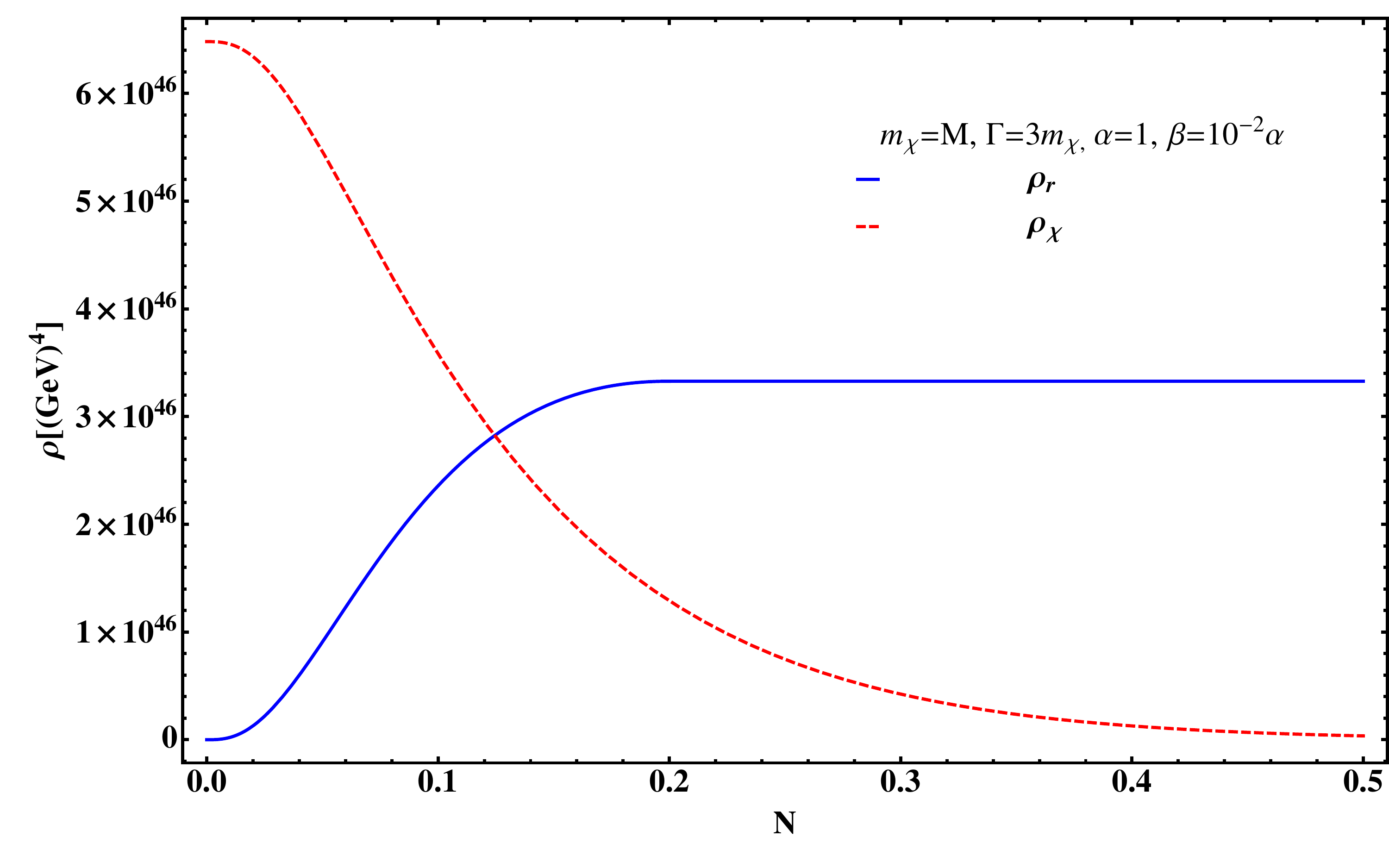,width=0.5\linewidth,clip=}
\end{tabular}
\vspace{-0.5cm}
\caption{The evolutions of $\rho_{r}$ and $\rho_{\chi}$ during the oscillatory epoch and the reheating epoch when $m_{\chi} = M$ and $\beta = 10^{-2} \alpha$. {\it left}) The reheating process is slow because of the small value of the decay rate, $\Gamma = 10^{-3} m_{\chi}$. {\it right} ) The reheating process is very efficient due to the large value of the decay rate, $\Gamma = 3 m_{\chi}$. } \label{fig-4}
\vspace{1cm}
\end{figure}

\begin{figure}[h]
\centering
\vspace{1cm}
\begin{tabular}{cc}
\epsfig{file=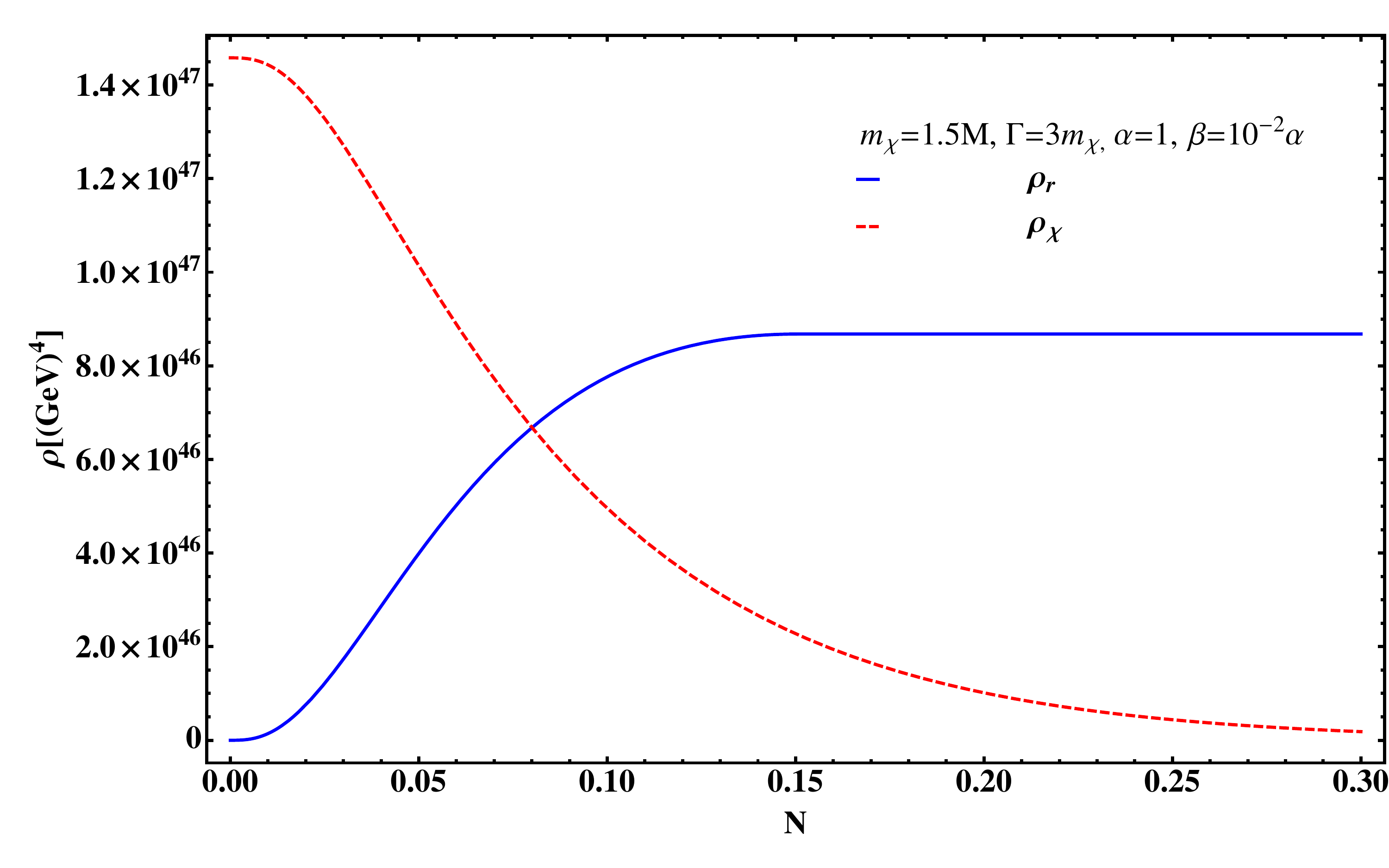,width=0.5\linewidth,clip=} &
\epsfig{file=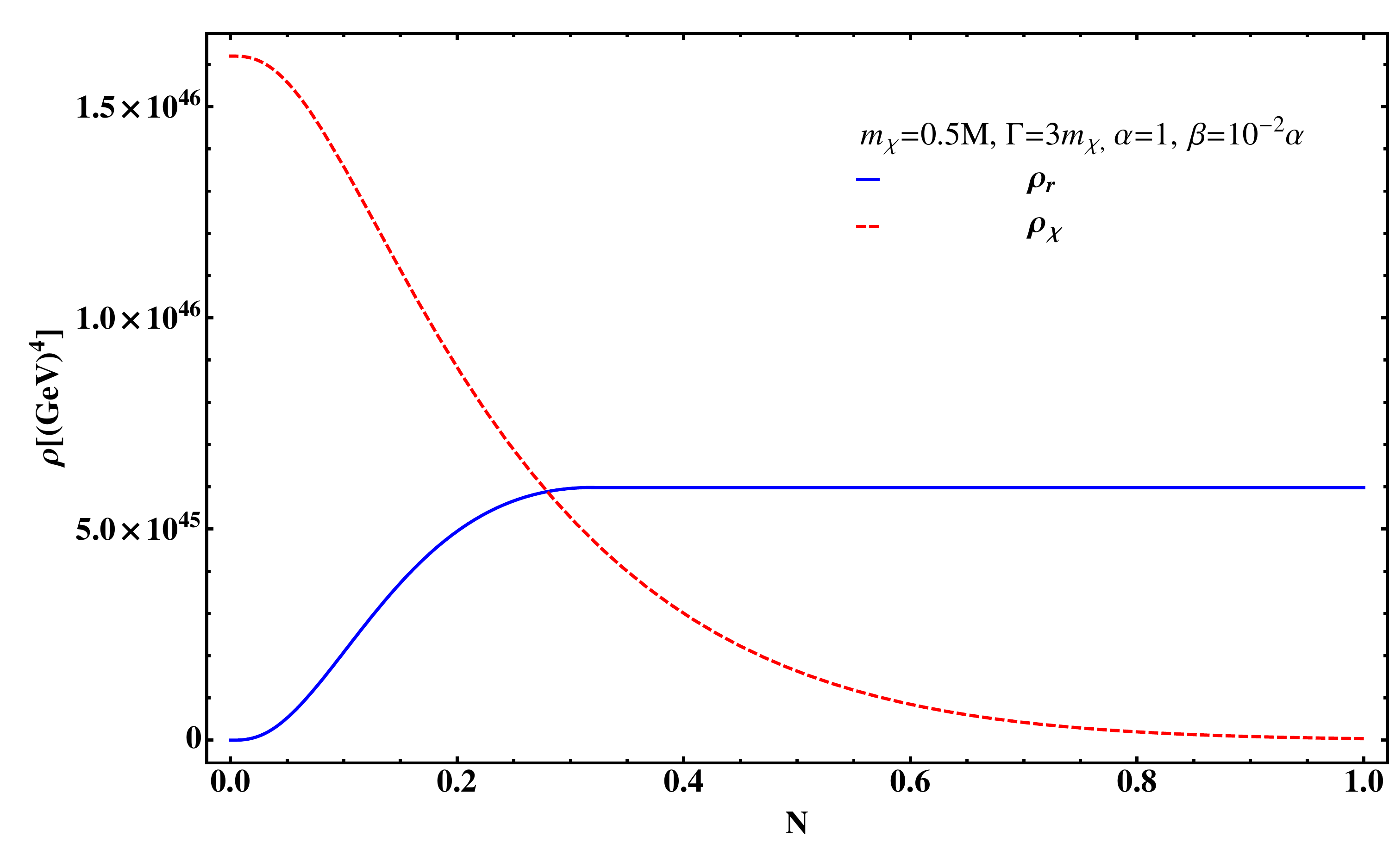,width=0.5\linewidth,clip=}
\end{tabular}
\vspace{-0.5cm}
\caption{The evolutions of $\rho_{r}$ and $\rho_{\chi}$ when $\Gamma = 3 m_{\chi}$ and $\beta = 10^{-2} \alpha$. {\it left}) The reheating temperature is higher due to the large scalar field mass, $m_{\chi} = 1.5M$. {\it right}) The reheating happens later due to the small value of scalar field mass, $m_{\chi} = 0.5M$. } \label{fig-5}
\vspace{1cm}
\end{figure}

The scalaron reaches to its minimum potential at $\phi = 0$ ({\it i.e.} $R=0$) and starts to oscillate. These oscillations excite the fields and reheat the Universe. To estimate the reheating, one can add the simple case of a scalar field $\chi$ to the original action in Eq.(\ref{Sfx})  
\begin{align} S &\equiv \int d^{4} x \sqrt{-g} \left( \frac{\bMp^2}{2} f \left(R \right) + {\cal L}_{\chi} + {\cal L}_{r} \right) \label{SfRchi}  \\
&= \frac{1}{2} \int d^{4} x \sqrt{-g} \left( \bMp^2 f \left(R \right) -  \tensor{g}{^\mu^\nu} \partial_{\mu} \chi \partial_{\nu} \chi - m_{\chi}^2 \chi^2 \right) + S_{r} \, , \nonumber \end{align}
where $S_{r}$ is the action of the radiation which is produced by the decay of $\chi$ field after the inflation. In this scenario, we do not consider the coupling between $R$ and $\chi$ ({\it i.e.} no decay of the scalaron). One can use the same field equation given in Eq.(\ref{fEq}) by specifying the energy momentum tensor as
\begin{equation} \tensor{T}{^\mu^\nu} \equiv {T_{\chi}}\tensor{\vphantom{T}}{^\mu^\nu} + {T_{r}}\tensor{\vphantom{T}}{^\mu^\nu}  = \partial^{\mu} \chi \partial^{\nu} \chi - \tensor{g}{^\mu^\nu} {\cal L}_{\chi} + \left( \rho_{r} + P_{r} \right) U^{\mu} U^{\nu} + P_{r} \tensor{g}{^\mu^\nu} \, , \label{Tmunu} \end{equation} 
where $U^{\mu}$ is the four-velocity and $\rho_{r}$ and $P_{r}$ are the energy density and pressure of the radiation, respectively. In the FRW metric, one obtains the field equations for the $\chi$ and radiation,
\begin{align} \ddot{\chi} + 3H \dot{\chi} + m_{\chi}^2 \chi &= -\Gamma_{\chi} \dot{\chi} \,\, , \label{chifeq} \\
\dot{\rho}_{r} + 4 H \rho_{r} &= \Gamma_{\chi} \dot{\chi}^{2} \,\, . \label{dotrho} \end{align} 
 Because one is interested in the oscillatory region in the background produced by the scalaron, the background evolutions are given by Eqs.(\ref{Htiosc})-(\ref{Rtiosc}). It means that one can ignore any backreaction on the background evolutions from $\chi$ and $\rho_{r}$ during this period. Thus, one can solve the field equations (\ref{chifeq}) and (\ref{dotrho}) both analytically and numerically. For this purpose, one can rewrite these equations as a function of e-folding numbers, $N$
 \begin{align}  \frac{d^2 \chi}{dN^2} + \left( \frac{3}{2} + \frac{\Gamma_{\chi}}{H\left(N\right)} \right) \frac{d \chi}{d N} + \frac{m_{\chi}^2}{H(N)^2} \chi &= 0 \,\, , \label{chifeqN} \\
\frac{\rho_{r}}{dN} + 4 \rho_{r} - \Gamma_{\chi} H(N) \left( \frac{d \chi}{d N}  \right)^{2}&= 0 \,\, , \label{Nrhor} \\
 \text{where} \,\, N \equiv \ln \left[ \frac{a}{a_{\osc}} \right] = \frac{2}{3} \ln \left[ 1 + \frac{\omega \left(t - \tosc \right)}{4} \right] \,\, , \,\,  H(N) &= \frac{\omega}{6} e^{-\frac{3}{2} N} \,\, .  \label{HN} \end{align} 
As can be seen in Eq.(\ref{Nrhor}), $\rho_{r}$ is created from the decaying of the $\chi$ field. At the end of the inflation, $\Gamma$ is typically much smaller than the Hubble parameter. Thus, at the beginning of the phase of the inflationary oscillations, the energy loss into particles is negligible compared to the energy loss due to the expansion of the space time. $\chi$ particle production becomes effective only when the Hubble expansion rate decreases to a value comparable to $\Gamma$. Thus, $\rho_{r}$ becomes the constant value of $\Gamma_{\chi} H(N_{\osc}) (d\chi/dN |_{N=N_{\osc}})^2$ at the not too long after the Universe has come into the oscillation phase.
 One can define the end of the reheating when the 90\% of the energy density of the $\chi$ field is converted to that of the radiation ({\it i.e.} $\rho_{r} = 9 \rho_{\chi}$) where $\rho_{\chi}$ is given by
 \begin{equation} \rho_{\chi} = \frac{1}{2} \left( H^2 \left( \frac{d \chi}{d N} \right)^2 + m_{\chi}^2 \chi^2 \right) = \frac{H^2 M^2}{2} \left( \left( \frac{d \bar{\chi}}{d N} \right)^2 + \frac{m_{\chi}^2}{H^2} \bar{\chi}^2 \right) \, , \label{rhochi} \end{equation}
 where we define the dimensionless field variable $\bar{\chi} = \chi /M$. The energy density of the radiation at the temperature $T$ is given by
 \begin{equation} \rho_{r}(N) = \frac{g_{\ast}(N) \pi^2}{30} T(N)^{4} \, , \label{rhor} \end{equation}
 where $g_{\ast}(N)$ is the number of relativistic species degrees of freedom and it is around ${\cal O}(100)$ at an early epoch. From the definition of the reheating epoch, one can obtain the reheating temperature 
\begin{equation} T_{\rh} = \left( \frac{9}{50 \pi^2} \left( \frac{100}{g_{\ast}} \right) \rho_{\chi}(N_{\rh}) \right)^{1/4} \, . \label{Trh} \end{equation}
The reheating temperature is determined by numerically solving Eqs.(\ref{chifeq}) and(\ref{dotrho}). This depends on model parameters, $\alpha$, $\beta$, $M$, $\Gamma_{\chi}$, and $m_{\chi}$. We demonstrate the evolutions of $\rho_{\chi}$ and $\rho_{r}$ for the different models in Figs.\ref{fig-4} and \ref{fig-5}. 

In the figure.\ref{fig-4}, we depict the evolutions of $\rho_{r}$ and $\rho_{\chi}$ when $m_{\chi} = M$ and $\beta = 10^{-2} \alpha$. In the left panel of Fig.\ref{fig-4}, we choose the small decay rate, $\Gamma = 10^{-3} m_{\chi}$. In this case, it takes longer time ({\it i.e.} larger $N_{\text{rh}}$) to reheat the Universe compared to the larger decay rate models. Thus, one obtains the lower reheating, $T_{\text{rh}} = 2.3 \times 10^{10}$ GeV. In this model, the $\rho_{r}$ reaches to its maximum values at $N_{\text{max}} = 0.37$ after the beginning of the oscillatory epoch and terminates the reheating at $N_{\text{rh}} = 3.77$. In the right panel of the figure.\ref{fig-4}, the decay rate is now $3 m_{\chi}$. Thus, reheating process becomes faster and after $\rho_{r}$ reaches its maximum at $N_{\text{max}} = 0.20$, it becomes thermalized at $N_{\text{th}} =  0.31$. In this model, the reheating temperature is higher than the first model and becomes $T_{\text{rh}} = 9.1 \times 10^{10}$ GeV.  

In the figure.\ref{fig-5}, the evolutions of $\rho_{r}$ and $\rho_{\chi}$ are depicted for different values of $m_{\chi}$ with $\Gamma = 3 m_{\chi}$ and $\beta = 10^{-2} \alpha$. In the left panel of Fig.\ref{fig-5}, the large mass scalar field, $m_{\chi} = 1.5 M$ is chosen. In this case, the decay process is more efficient than the smaller mass model and it takes less time ({\it i.e.}  $N_{\text{rh}}$) to reach both the maximum $\rho_{r}$ and the reheating of the Universe compared to the smaller mass of the scalar field models. Thus, one obtains the higher reheating, $T_{\text{rh}} = 1.2 \times 10^{11}$ GeV. In this model, the $\rho_{r}$ reaches to its maximum values at $N_{\text{max}} = 0.15$ after the beginning of the oscillatory epoch and terminates the reheating at $N_{\text{rh}} = 0.21$. In the right panel of the figure.\ref{fig-5}, the mass of the scalar field, $m_{\chi} = 0.5 M$. Thus, reheating process becomes slower and after $\rho_{r}$ reaches its maximum at $N_{\text{max}} = 0.32$, it becomes thermalized at $N_{\text{th}} =  0.64$. In this model, the reheating temperature is lower than the model of the left panel, $T_{\text{rh}} = 5.9 \times 10^{10}$ GeV.  

\begin{table}
\centering
\caption{The reheating temperature for the different models when $M = 6 \times 10^{12}$ GeV. We assume that the reheating phase is obtained when $\rho_{\chi}$ reaches to 90 \% of the maximum $\rho_{r}$.}
\label{tab-3}
\vspace{0.5cm}
\begin{tabular}{|c|c|c|c|c|c|c|c|}
\hline
\multirow{2}{*}{$m_{\chi }$} & \multirow{2}{*}{$\Gamma$}  & \multirow{2}{*}{$\beta$}  & $\rho _r\left(N_{\max }\right)$ & \multirow{2}{*}{$N_{\max }$} & $\rho _{\chi }\left(N_{\text{rh}}\right)$ & \multirow{2}{*}{$N_{\text{rh}}$} & $T_{\text{rh}}$ \\ 
		      &                        &                   & $\left[(\text{GeV})^4\right]$ &  & $\left[(\text{GeV})^4\right]$ &  & $[\text{GeV}]$ \\ [1ex]
\hline
\multirow{2}{*}{M} & \multirow{2}{*}{$10^{-3}m_{\chi }$} & $-10^{-2}$ & $1.33\times 10^{44}$ & $0.37$ & $1.49\times 10^{43}$ & $3.77$ & $2.28\times 10^{10}$ \\ [1ex]
     &                                    & $10^{-2}$  & $1.30\times 10^{44}$ & $0.37$ & $1.48\times 10^{43}$ & $3.77$ & $2.28\times 10^{10}$ \\ [0.5ex]
\hline
 M & $3m_{\chi }$ & $10^{-2}$ & $3.33\times 10^{46}$ & 0.20 & $3.70\times 10^{45}$ & 0.31 & $9.06\times 10^{10}$ \\ [0.5ex]
\hline
 1.5M & $3m_{\chi }$ &$10^{-2}$& $8.68\times 10^{46}$ & 0.15 & $9.64\times 10^{45}$ & 0.21 & $1.15\times 10^{11}$ \\ [0.5ex]
\hline
 0.5M & $3m_{\chi }$ & $10^{-2}$ & $5.98\times 10^{45}$ & 0.32 & $6.64\times 10^{44}$ & 0.64 & $5.90\times 10^{10}$ \\ [0.5ex]
\hline
\end{tabular}
\end{table}
We summarize the results in the table.\ref{tab-3}. As shown in the figures.\ref{fig-4} and \ref{fig-5}, the larger the decay rate, the earlier the reheating epoch.  Thus, for the same values of $m_{\chi}$ and $\beta$, one obtains the higher reheating temperature for the larger decay rate. Also, if $m_{\chi}$ increases, so does the $T_{\text{rh}}$ due to the increasing of efficiency of the decay of the scalar field. We are also interested in the effect of the $\beta$ on the reheating temperature in order to investigate the possible constraint on its values from the reheating temperature. Even though, there exists small deviations on the reheating temperatures for the different values of $\beta$, the difference is less than 1 \%.

\section{Conclusions}
We have investigated the general modification of the Starobinsky inflation model by including the logarithmic correction in addition to the quadratic in the Ricci scalar. We show that the values of coefficients $\alpha$ and $\beta$ are strongly constrained from the observation. The maximum value of the coefficient $\beta$ is about 1 \% of that of the coefficient $\alpha$ in order to be satisfied with the CMB results. $\beta \neq 0$ means the deviation of the model from the Starobinsky's one. When we vary the value of $\beta$ from 0 to |0.02| ({\it i.e.} 2 \% change in $\beta$), the scalar spectral index is changed by 1\% only. We also investigate the reheating process in order to obtain any deviation from the Starobinsky model. The changes in the reheating temperatures are less than 1 \% when we compare all the viable $\beta$ values. Even though, the effects of the deviation from the Starobinsky model on the known observational quantities are quite small, one might still be able to confirm the models with upcoming more accurate observations and (or) from other cosmological observables.       

\begin{appendices}
In this appendix, we show the detail derivation of the damped oscillation behavior given in the Eq.(\ref{Hosc}).  

\section{Oscillation}
\renewcommand{\theequation}{\thesection.\arabic{equation}}
\setcounter{equation}{0}

If one uses the Eq.(\ref{bphi}), then Eq.(\ref{bphiEq}) is given by
\begin{equation}  \label{ffieldEq} \end{equation} 

One can obtain the approximate damped oscillation of the Hubble parameter by replacing $c_{1}$ by function $g(t)$ in the equation (\ref{Hosc})
\begin{equation} H_{\text{osc}} (t) \simeq g(t) \times \cos^{2} \left[\omega \left(t - \tosc \right) \right] \,\, , \,\, \tosc \leq t \, , \label{Hoscdampapp} \end{equation}
where $\omega^2 = M^2/(24 \alpha)$. In order to obtain the proper damped oscillation, the form of $g(t)$ should be given by 
\begin{equation} g(t) = \left(  c_{1(\osc)} + c_{2(\osc)} \left( t - \tosc \right) + c_{3(\osc)} \cos \left[ 2 \omega \left( t - \tosc \right) \right] \right)^{-1}  \, , \label{gtapp} \end{equation}
where $c_{i(\osc)}$ are constant to be determined from the boundary values of $H(\tosc)$ before and during the oscillation period. The boundary conditions are given by
\begin{align} H(\tosc) &= H_{i} - \frac{2}{3} \omega^2 \left( \tosc - t_{i} \right) = H_{\osc} \left(\tosc \right) = \frac{1}{c_{1(\osc)} + c_{3(\osc)}} \, , \label{BC1} \\ 
\dot{H}(\tosc) &= -\frac{2}{3} \omega^2 = \dot{H}_{\osc} \left(\tosc \right) = - \frac{c_{2(\osc)}}{\left( c_{1(\osc)} + c_{3(\osc)}\right)^2} \, , \label{BC2} \\\ddot{H}(\tosc) &= 0 = \ddot{H}_{\osc} \left(\tosc \right) = \frac{2\left( c_{2(\osc)}^{2} +\left( c_{3(\osc)}^2 - c_{1(\osc)}^2 \right) \omega^2 \right)}{\left( c_{1(\osc)}  + c_{3(\osc)} \right)^3} \, , \label{BC3}  \end{align}
From these equations.(\ref{BC1})-(\ref{BC3}), one obtains
\begin{equation} c_{1(\osc)}  = \frac{15}{2\omega} \, , \, c_{2(\osc)}  = 6 \, , \, c_{3(\osc)} = - \frac{9}{2\omega} \, . \label{c1c2c3} \end{equation}
Thus, one obtains the analytic solution of $H(t)$ during the oscillation period as
\begin{equation} H_{\osc} (t) = \left(\frac{15}{2\omega }+6(t-\text{tosc})-\frac{9}{2\omega } \cos [2\omega (t-\tosc)]\right)^{-1} \cos^{2} \left[\omega \left(t - \tosc \right) \right] \, . \label{Hosc3} \end{equation}

energy density of the radiation
\begin{align} \rho_{r}(N_{\osc}) &\simeq \Gamma_{\chi} H(N_{\osc}) M^2 \chi'\left(N_{\osc}\right)^2 \, , \label{rhoNosc} \\
\rho_{\chi}(N_{\osc}) &\simeq \frac{H(N_{\osc})^2 M^2}{2} \left( \bar{\chi}'(N_{\osc})^{2} + \frac{m_{\chi}^2}{H(N_{\osc})^2} \bar{\chi}(N_{\osc})^2 \right) \, ,\label{rhochiNosc} \\
\rho_{\chi}(N_{\rh}) &= \frac{H(N_{\rh})^2 M^2}{2} \left( \bar{\chi}'(N_{\rh})^{2} + \frac{m_{\chi}^2}{H(N_{\rh})^2} \bar{\chi}(N_{\rh})^2 \right) \, , \label{rhochiNrh} \\
\rho_{r}(N_{\rh}) &= \frac{\pi^2}{2} g_{\ast} T_{\rh}^4 \equiv 9 \rho_{\chi}(N_{\rh}) \, . \label{Trh} \end{align}
Thus, one can obtain reheating temperature $T_{\rh}$ as
\begin{equation} T_{\rh} = \left( \frac{9}{50 \pi^2} \left( \frac{100}{g_{\ast}} \right) \rho_{\chi}(N_{\rh}) \right)^{1/4} \, . \label{Trh2} \end{equation}

\end{appendices}

\section{Acknowledgments} 
SL is supported by Basic Science Research Program through the National Research Foundation of Korea (NRF)
funded by the Ministry of Science, ICT and Future Planning (Grant No. NRF-2017R1A2B4011168).

\end{document}